\documentstyle[seceq,preprint]{ptptex}

\def\beqra{\begin{eqnarray}}
\def\eeqra{\end{eqnarray}}
\def\beqast{\begin{eqnarray*}}
\def\eeqast{\end{eqnarray*}}
\def\be{\begin{enumerate}}
\def\ee{\end{enumerate}}

\def\beq{\begin{equation}}
\def\eeq{\end{equation}}

\title{
See-saw Mechanism and Possible Generation Structure
}

\author{
Kazuo {\sc Koike}
}

\inst{
Department of Natural Science,~Kagawa Univrsity,

Takamatsu 7608522
\\
}

\recdate{
}

\abst{
 
  On the basis of see-saw mechanism, possible generation structure is
investigated. Assuming that right-handed neutrinos are realistic 
particles below GUTs mass, it is shown that there is a choice of new 
scheme of generation,
where the combination of leptons and quarks is different from the
ordinary one. Our scheme predict a new proton decay mode.

}

\begin{document}

\maketitle

\section{Introduction}

 The problem of generation structure which is observed in low energy 
region seems to exhibit an important suggestion to disclose  fundamental 
existence-form of matter.\cite{rf:Maki,rf:KK} In investigating to that 
structure, neutrino problem may be situated at the capstone. It seems 
that the see-saw mechanism is the most prominent one to explain the 
smallness of neutrino mass, where its smallness is reduced to 
largeness of the right-handed Mayorana neutrino mass which is often 
assumed to be the order of GUTs mass. In the GUTs scale, a new 
physics beyond to  our present prediction may appear,
and it is not clear whether the right-handed Mayorana neutrino with
GUTs mass can be treated as ordinary particle.
 On the other hand, if we assume that the right-handed Mayorana 
neutrino possesses mass 
below GUTs scale, and can be completely treated as ordinary particles
described by ordinary gauge field theory,
possible alternative generation structure will be implied.
 
In this paper, we will investigate possible new generation structure
on the basis of viewpoint suggested by the see-saw mechanism, and 
discuss on a few typical new features due to this scheme.



\section{See-saw mechanism of neutrino mass splitting}

\indent
 For definiteness, we will make a quick review of see-saw mechanism of
neutrino mass generation.\cite{rf:GRSY}
As a basis of construction of our scheme, let
us consider the D-M (Dirac-Mayorana) mass term 
\cite{rf:Pon}\tocite{rf:Bilen} in the simplest case of
one generation labeled by the generation suffix i.
We have

\begin{eqnarray}
{\cal{L}}^{D-M}& = &
-\frac{1}{2} m_{iL} \overline{(\nu_{iL})^c} \nu_{iL}
-m_{iD}\bar{\nu}_{iR}\nu_{iL}
-\frac{1}{2} m_{iR} \bar{\nu}_{iR} (\nu_{iR})^c~~+~~h.c.
                  \nonumber\\
               & = &-\frac{1}{2}
\overline{{(\nu_{iL})^c} \choose {\nu_{iR}}}
{}~M~{{\nu_{iL}} \choose {\nu_{iR}^c}}
{}~~+~~h.c.
\label{eq:DM}
\end{eqnarray}

\noindent Here

\beq
M = \left( \begin{array}{cc}
m_{iL} & m_{iD} \\
m_{iD} & m_{iR}
\end{array} \right),
\label{eq:MATm}
\eeq

\noindent $m_{iL}, m_{iD}, m_{iR} $ are parameters. For a symmetrical 
matrix M we have

\beq
M=U~ m~ U^{\dag},
\label{eq:DIA}
\eeq

\noindent where $U^{\dag}U=1 $, $m_{jk}=m_j \delta_{jk} $. From 
Eq.(\ref{eq:DM}) and
Eq.(\ref{eq:DIA}) we have

\beq
{\cal{L}}^{D-M}=-\frac{1}{2} \sum_{\alpha=1}^{2}
 m_{i{\alpha}}{\bar{\chi}}_{{i\alpha}} \chi_{i{\alpha}} ,
\label{eq:dia2}
\eeq

\noindent where

\begin{eqnarray}
\nu_{iL}~~  =~~~{\cos{\theta_i}}\chi_{i1L} & + & {\sin{\theta_i}}\chi_{i2L}
\nonumber\\
(\nu_{iR})^c  =  {-\sin{\theta_i}}\chi_{i1L} & + & {\cos{\theta_i}}\chi_{i2L}.
\label{eq:MIX}
\end{eqnarray}

\noindent Here $\chi_{i1} $ and $\chi_{i2} $ are fields of Majorana
 neutrinos with masses $ m_{is}~(small), m_{iB}~(Big) $, respectively.
  The masses $m_{is}, m_{iB} $ and the mixing angle
$\theta_i $ are connected to the parameters $m_{iL}, m_{iD} $ 
and $m_{iR} $ by the relations

\begin{eqnarray}
m_{is}~ & = & \frac{1}{2}~ {\left|{m_{iR}~ +~m_{iL}~-~a_i} \right|}
\nonumber\\
m_{iB} & = & \frac{1}{2}~ {\left|{m_{iR}~ +~m_{iL}~+~a_i} \right|}
\nonumber\\
\sin{2\theta_i} & = & \frac{2m_{iD}}{a_i},~~~~\cos{2\theta_i}
=\frac{m_{iR}-m_{iL}}{a_i}
\label{eq:MIX2}
\end{eqnarray}

\noindent where

\beq
a_i=\sqrt{(m_{iR}-m_{iL})^2~+~4{m_{iD}^2}}
\label{eq:a}
\eeq

It should be noted that the relations Eq.(\ref{eq:MIX2}) are exact ones. 
Let us assume now that

\beq
m_{iL}=0,~ m_{iD}\simeq{ m_{iF}},~m_{iR}\gg{ m_{iF}} ,
\label{eq:mass-sb1}
\eeq

\noindent where $m_{iF} $ is the typical mass of the leptons and quarks of
the generation labeled by suffix i. From Eq.(\ref{eq:MIX2})  
we have

\beq
m_{is}\simeq\frac{m_{iF}^2}{m_{iR}},~~m_{iB}\simeq{m_{iR}} ,
{}~~\theta_i\simeq{\frac{m_{iD}}{m_{iR}}}
\label{eq:mass-sb2}
\eeq

\noindent
 Thus, if the conditions Eq.(\ref{eq:mass-sb1})  are satisfied,
the particles with definite masses are split to a very light Majorana
neutrino with mass
$m_{is} \ll m_{iF} $ and a very  heavy Majorana particle with mass
$m_{iB}\simeq m_{iR} $. The current neutrino field $\nu_{iL} $ practically
coincides with
 $\chi_{i1L} $ and $ \chi_{i2}\simeq {\nu_{iR}~+~(\nu_{iR})^c} $
, because $\theta_i$ is extremely small.

That is , we have assumed such scheme that in
D-M mass term Dirac masses  are of order of usual fermion masses, 
the right-handed Majorana masses, responsible for lepton numbers
violation, are  extremely large  and the left-handed 
Majorana masses are equal zero. In such a scheme  neutrinos are Majorana
particles with masses much smaller than masses of the other fermions.
The predictions of neutrino masses depend on the value of the
$m_{iR}$ mass.
The value of $m_{iR}$ is often assumed that 
$m_{iR}=M_{GUT}$, where $M_{GUT} $ is grand unification scale.
Though this value  depends on the model, a typical one is
$m_R\simeq {10^{19}} $ GeV (Planck mass). In the $M_{GUT}$ region, 
the ordinary particle picture may be drastically changed.
\par
It should be emphasized that, however, there is no definite reason 
why the mass of  $m_{iR}$ should be $M_{GUT}$.  It is also possible 
that though the mass of  $m_{iR}$
is very huge it is  below the Planch mass and  possessing the picture of
ordinary particle. In such case,  possibility of new generation 
structure will be implied, which is discussed in the next section.

\section{Possible generation structure based on see-saw mechanism }

  The "standard" structure of generation is composed of each leptons 
and quarks, which generation number is labeled according to the sequence 
of their historical discovery, depending on the energy frontier. 
It should be emphasized that, in the present stage, the only compelling 
reason to relate leptons and quarks is the anomaly free condition, and no 
essential principle to compose the generation is ever unknown.\cite{rf:GUTs} 
Then, regarding the conventional rule to classify the generation according to 
the sequence of magnitude of the total masses belonging to the same suffix as 
the fundamental rule which should be founded on the deeper level, we will 
investigate alternative possible generation structure.

If the right-handed Majorana neutrinos are realistic particles below the 
GUT mass and responsible to the see-saw mechanism, its existence will 
affect to the above mentioned classification procedure of the generations.
From Eq.(\ref{eq:mass-sb2}),
\beq
m_{iR}\simeq\frac{m_{iF}^2}{m_{is}}.
\label{eq:mass-sb3}
\eeq
Remembering to the fact that the current neutrino field $\nu_{iL} $ 
practically coincides with  $\chi_{i1L} $ with the mass $m_{is}$
together with the above relation 
Eq.(\ref{eq:mass-sb3}),
let us investigate the new generation scheme with realistic $m_{iR}$. 
Though the values of neutrino mass has not yet been established,
\cite{rf:nu-mix} a prominent
possibility is that the mass of neutrinos exhibits an extreme hierarical
structure\cite{rf:Tanimo},
\beq
m_{\nu_e}\ll{m_{\nu_\mu}}\ll{m_{\nu_\tau}}.
\label{eq:mass-inequal}
\eeq
The degree of the mass difference in Eq.(\ref{eq:mass-inequal})
will read to a suggestion to find possible new generation structure.
That is, if the degree of mass difference of neutrinos is larger than that of
the representative mass of each generation $m_{iF}$, the magnitude of total 
mass of i-th neutrinos,
i.e. $m_{\nu_i} + m_{iR}$, will become in  reverse order to label number i,
and this magnitude will dominate the total lepton masses belonging to i-th 
label.
Then, provided that we adopt the custom to nominate the generation number 
"from light to heavy" particles, we will conclude  the following generation 
classification, 


\beq
\begin{array}{lll}
\\
I&~~(\nu_{1R})^c~~~{{\nu_{\tau}} \choose {\tau}}~~& {{u_L} \choose {d_L}}\\
\\
II&~~(\nu_{2R})^c~~~{{\nu_{\mu}} \choose {\mu}}~~ &{{c_L} \choose {s_L}}\\
\\
III&~~(\nu_{3R})^c~~~{{\nu_{e}} \choose {e}}~~ &{{t_L} \choose {b_L}}\\\
\end{array}
\label{eq:newgene}
\eeq
where we have written only the left-handed components.
That is, taking into account of the existence of realistic particles $\nu_{iR}$,
our proposal for new generation structure is different from the ordinary
one. The characteristic feature is that the leptons belonging to the first and 
the third generations are exchanged. It should be noted that this result is 
caused by the existence of realistic $\nu_{iR}$ and the see-saw mechanism.

\section{"Generational" and "Inter-generational" gauge bosons in GUTs
and proton decay problem}

The generation structure is the important feature
observed in low energy region. We will suppose that this structure still 
leaves any trace  in  high-energy region below GUTs scale. This 
situation leads to the viewpoint to distinguish the gauge bosons appearing
in GUTs containing all generations. That is,
the gauge bosons common to all generations, and ones which connect
particles belonging to different generations.
In SU(5) GUT, only the 
former type of gauge bosons, $W, Z, A$ and $X, Y$ , appear. We will 
call this type of gauge bosons 
as "generational gauge bosons"\cite{rf:Abe-lq2}. In the GUTs structure 
containing all generations, new gauge bosons connecting particles
belonging to the different generations appear generally, and we will call
them as "inter-generational gauge bosons".\cite{rf:flavor-c} 
Our viewpoint means that the 
contribution of the "inter-generational gauge boson" should be more 
suppressed than that of the "generational gauge bosons" in low and 
intermediate energy region below the GUTs scale.

If we take this viewpoint, our model of new generation structure 
lead to different feature of proton decay problem in GUTs.
Our scheme predict the proton decay mode due to the generational gauge 
bosons $X$ and $Y$

\beq
P~\rightarrow~\tau^+~M_0
\label{eq:decay-mode}
\eeq

\noindent
instead of the well known mode

\beq
P~\rightarrow~e^+~M_0
\label{eq:decay-mode-e}
\eeq

\noindent
where $M_0$ represents $\pi^0,\rho_0,\omega,\eta,\pi^+\pi^-\cdots$.
It should be noted that the process in Eq.(\ref{eq:decay-mode}) is
forbidden by Q value.
The inter-generational gauge bosons may cause the process in 
Eq.(\ref{eq:decay-mode-e}), however, this process will be extremely
suppressed in low and intermediate energy region below GUTs scale.
 The other mode is

\beq
P~\rightarrow~{\overline{\nu}}_\tau~M^+
\label{eq:decay-mode-nu}
\eeq

\noindent
with $\pi^+,\rho^+,\pi^+\pi^0\cdots$, and this process is allowed
in our scheme.

Thus, the proton decay in our model is different from well known
model of generation structure. That is, the well-known difficulty 
so-called "proton decay problem" is consistent to our scheme. It 
should be emphasized that when the proton decay is actually observed 
it will be made clear whether our scheme is realized or not.  

\section{Discussion}

  In this paper, we have proposed a possible new structure of generation 
combination of leptons and quarks on the basis of see-saw mechanism
and the actual existence of realistic right-handed Mayorana neutrino. 
We have  emphasized  classification of the gauge bosons appearing in GUTs, 
that is the "generational gauge bosons" and the "inter-generational gauge 
bosons". Our model gives some important predictions to so-called "proton 
decay problem". In order to give further predictions, some specific assumptions 
and introduction to many parameters is inevitable in the present stage, 
then in this paper we have restricted ourselves to proposal of 
framework of the model.

 We have discussed on only one generation case in order to investigate the 
fundamental framework of new possibility. However, massive neutrinos will 
generally cause the generation mixing.[Appendix] In fact, it is pointed out 
that the large mixing will be realized in the atomosphear neutrino events.
\cite{rf:SK-Atomospher} In the framework of ordinary generation structure, 
it seems that new mysterious feature may appear provided that we 
take into account of this large mixing in GUTs mass relation.
That is , the generation labels of  leptons seem to be exchanged in 
appearance between the first and the third generation.\cite{rf:Band} 
It should be emphasize that, in our model this appearance is not the 
exchange but the fundamental generation structure itself. Further 
analysis concerning to the problem of generation mixing will be 
discussed in elsewhere. 

 Finally, it should be noted that our discussion to introduce the generation 
combination may be heuristic one based on the historical or conventional 
rule of generation nomination in the present stage, then our model will 
maintain its meaning even if some premises are partially altered. This simple 
rule will be based on the deeper level structure of nature, and will be 
justified in near future.

\vfill\eject







\appendix
\section{}
 In the general case of 3-generations, the mixing has, instead of 
 Eq.(\ref{eq:DM}),the following form,

\begin{eqnarray}
{\cal{L}}_{D-M}&=&-~\frac{1}{2}
\sum_{l,l'} \overline{(\nu_{l'L})^c} M_{l'l}^L \nu_{lL}~
-\sum_{l',l} \bar{\nu}_{l'R} M_{l'l}^{D} \nu_{lL}~
                                  \nonumber\\
               & &-~\frac{1}{2}\sum_{l',l}\bar{\nu}_{l'R} M_{l'l}^{R}
(\nu_{l'R})^c~~+~~h.c.,
\label{eq:Mixgeneral}
\end{eqnarray}

\noindent where $M^L, M^D $ and $M^R $ are 3 $\times $ 3
complex matrixes.

It is clear that,  neutrinos with definite masses are Majorana particles.
As in the case of one generation given in Eq.(\ref{eq:dia2}), the number 
of massive particles in this case is twice of the
number of lepton flavours.From Eq.(\ref{eq:Mixgeneral}), after 
standard procedure of the diagonalization of a 6 $\times $6 matrix M 
we have

\beq
{\cal{L}}_{D-M}=-\frac{1}{2} \sum_{\alpha=1}^6 m_{\alpha}{ \bar{\chi}_\alpha} 
\chi_\alpha ,
\label{eq:Summation6}
\eeq

\noindent where  $\chi_\alpha=\chi_\alpha^c $  is the field of Majorana 
neutrinos
with mass $m_\alpha $. The current fields $\nu_{lL} $ and  fields
$(\nu_{lR})^c=C \bar{\nu}_{lR}^T $~ (left-handed components) are
connected with left-handed components of massive Majorana fields
$\chi_{iL} $ by a unitary transformation

\begin{eqnarray}
\nu_{lL}=\sum_{\alpha=1}^6 U_{l\alpha} \chi_{\alpha L}
\nonumber\\
(\nu_{lR})^c=\sum_{\alpha=1}^6 U_{\bar{l}\alpha} \chi_{\alpha L}
\label{eq:Mix6}
\end{eqnarray}

\noindent where U is a unitary 6 $\times $ 6 matrix.

If all masses $m_\alpha $ are small enough, the
fields $\nu_{lL} $ are usual flavor left-handed neutrinos and
right-handed antineutrinos, while the fields $\nu_{lR} $
are "sterile" right-handed neutrinos and left-handed antineutrinos,
and do not take place in the standard weak interaction.

\vfill\eject



\begin{thebibliography}{99}
\bibitem{rf:GUTs}
The mass relation in GUTs means 
merely indirect condition. If new structure of generation is established,
the modified GUTs relation will be constructed on the basis of it.
\bibitem{rf:nu-mix}
  In order to determine the neutrino masses, the mixing structure should
be completely clarified. In the present stage, it will be premature to find
the definite masses of neutrinos.
\bibitem{rf:flavor-c}
 "Inter-generational gauge bosons" will generally cause 
 flavor-changinng effects.
\end{thebibliography}
\end{document}